\documentclass[prd,amsmath,amssymb,twocolumn]{revtex4}

\def\de{\delta}

\def\th{\theta}

\def\ka{\kappa}

\def\De{\Delta}

\def\Om{\Omega}

\def\prt{\partial}

\def\fr#1#2{{{#1} \over {#2}}}
\def\half{{\textstyle{1\over 2}}}
\def\lsim{\mathrel{\rlap{\lower4pt\hbox{\hskip1pt$\sim$}}
    \raise1pt\hbox{$<$}}}
\def\gsim{\mathrel{\rlap{\lower4pt\hbox{\hskip1pt$\sim$}}
    \raise1pt\hbox{$>$}}}
\def\sqr#1#2{{\vcenter{\vbox{\hrule height.#2pt
         \hbox{\vrule width.#2pt height#1pt \kern#1pt
         \vrule width.#2pt}
         \hrule height.#2pt}}}}
\newcommand{\beq}{\begin{equation}}
\newcommand{\eeq}{\end{equation}}
\newcommand{\bea}{\begin{eqnarray}}
\newcommand{\eea}{\end{eqnarray}}
\newcommand{\rf}[1]{(\ref{#1})}
\newcommand{\nn}{\nonumber}

\def\etal{{\it et al.}}

\def\re{{\rm Re}}
\def\im{{\rm Im}}
\def\phat{\hat{\boldsymbol p}}
\def\pvec{\boldsymbol p}
\def\pmag{|\boldsymbol p|}
\def\kvec{\boldsymbol k}
\def\kavec{\boldsymbol\ka}
\def\ring#1{{\mathaccent'27 #1}}
\def\cri{{\ring{c}}}
\def\ccfc#1#2{\cri^{(#1)}_{#2}}
\def\nuTemplate#1#2#3#4{\big(#1^{(#2)}\big)_{#3}^{#4}}
\def\cof#1#2{\nuTemplate{c_\text{of}}{#1}{#2}{}}
\def\aof#1#2{\nuTemplate{a_\text{of}}{#1}{#2}{}}

\def\pmags#1{|\boldsymbol p_{#1}|}

\def\cv{\v Cerenkov} 

\begin{document}

\title{Testing Relativity with High-Energy Astrophysical Neutrinos}

\author{Jorge S.\ D\'\i az,$^1$ V.\ Alan Kosteleck\'y,$^1$ 
and Matthew Mewes$^2$}

\affiliation{$^1$Physics Department, Indiana University, 
Bloomington, Indiana 47405, USA\\
$^2$Physics Department, Swarthmore College,
Swarthmore, Pennsylvania 19081, USA}

\date{IUHET 575, August 2013}

\begin{abstract}

The recent observation of high-energy astrophysical neutrinos
can be used to constrain violations of Lorentz invariance 
emerging from a quantum theory of gravity.
We perform threshold and \cv\ analyses that improve 
existing bounds by factors ranging from about a million to $10^{20}$.

\end{abstract}

\maketitle

Astrophysical neutrinos offer exciting prospects
for a distinctive perspective on the Universe,
complementing existing photon and cosmic-ray techniques.
In addition to enhancing studies of astrophysical objects,
the detection of high-energy neutrinos from remote sources
also provides opportunities 
for unique studies of physics at the fundamental level.
In this work,
we use the recent observations of TeV-PeV neutrinos
in the IceCube detector at the South Pole
and the evidence adduced in favor of their astrophysical origin 
\cite{icecube1,icecube2}
to perform sensitive tests of Lorentz invariance,
the foundational symmetry of relativity.
Tiny observable violations of Lorentz invariance 
arising from new physics at the unification scale
are proposed features of some underlying theories 
combining the fundamental interactions
such as strings
\cite{ksp},
but detecting these effects is expected to be challenging
due to their likely suppression by factors
involving the ratio of a light scale to the Planck scale. 
The advent of high-energy neutrino astrophysics
vastly extends the prospects for tests with Planck-scale sensitivity
because the combination of the long baseline and the high energy
offers unique access to Planck-suppressed effects.

Neutrinos have several features making them well suited
to studies of Lorentz invariance.
The interferometric nature of neutrino oscillations 
makes them exquisitely sensitive to
certain kinds of relativity violations,
and several oscillation experiments 
have already achieved Planck-scale sensitivity 
\cite{nuexpt}.
In addition,
the tiny or zero neutrino mass and weak neutrino interactions
permit high sensitivity to types of Lorentz violation
that are inaccessible to oscillation experiments
\cite{km12,dkl}.
For example,
time-of-flight experiments 
comparing neutrino propagation against other particle species
and studies of neutrino interactions and decay processes
offer options for relativity tests
that are unique and complementary to oscillation searches.

The general framework for studying Lorentz violation
is the Standard-Model Extension (SME)
\cite{sme},
which is an effective field theory
constructed from General Relativity and the Standard Model
containing all operators for Lorentz violation.
Each operator is controlled by a coefficient for Lorentz violation,
and all operators describing neutrino propagation
have recently been classified and enumerated
\cite{km12}.
For the purposes of this work,
we can neglect oscillations of high-energy astrophysical neutrinos 
and can therefore work
in the context of the general oscillation-free model
or, where apposite, its isotropic limit.
Also,
since the observed high-energy IceCube events 
could be neutrinos or antineutrinos,
focusing on CPT-even operators is appropriate.
Here,
we obtain constraints
on oscillation-free coefficients for CPT-even Lorentz violation
deduced from threshold effects in hadron decays
and from \cv\ radiation.
Except where stated otherwise,
significant Lorentz violation 
is assumed to be limited to the neutrino sector,
compatible with the SME framework 
and current experimental bounds 
\cite{tables}.
The constraints reported in this work improve existing limits 
by factors ranging from $10^6$ to $10^{20}$.

In the relativistic, oscillation-free, and CPT-even limit,
the dispersion relation for a high-energy neutrino or antineutrino
of energy $E$ and momentum $\pvec$ is 
\cite{km12}
\beq
E(\pvec) = \pmag - \sum_{djm} \pmag^{d-3} Y_{jm}(\phat) \cof{d}{jm} ,
\label{dr}
\eeq
where $d=4,6,8,\ldots$ is the mass dimension of the underlying
operator in the field-theoretic action,
$j,m$ are conventional angular-momentum indices
with $0\leq j\leq d-2$,
and $\cof{d}{jm}$ are oscillation-free coefficients for Lorentz violation.
This expression omits the usual mass term,
which decreases with energy and can be neglected at high energies,
along with contributions from the odd-$d$ coefficients $\aof{d}{jm}$
associated with CPT-odd operators in the SME,
which appear with opposite signs for neutrinos and antineutrinos.
The Lorentz-violating modifications in Eq.\ \rf{dr}
introduce unconventional energy dependence 
as well as dependence on the propagation direction.
For each operator dimension $d$,
there are $(d-1)^2$ independent coefficients $\cof{d}{jm}$,
giving nine independent constant observables 
at the minimal dimension $d=4$
and 25 observables at the next order $d=6$.

Commuting boost generators produces rotations,
so every dispersion relation incorporating Lorentz violation 
is necessarily accompanied by some type of direction dependence.
However,
for situations where available data are insufficient
to perform a complete analysis of coefficients for Lorentz violation
at a given dimension $d$,
it is sometimes convenient to work within the isotropic limit
of the dispersion relation \rf{dr} 
to achieve order-of-magnitude estimates of the maximal potential sensitivity.
In this isotropic limit,
which holds only in a special frame,
the dispersion relation \rf{dr} reduces to
\cite{km12}
\beq
E(\pvec) = \pmag - \sum_{d} \pmag^{d-3} \ccfc{d}{} ,
\label{iso}
\eeq
where the isotropic coefficients $\ccfc{d}{}$ are defined by
$\ccfc{d}{} \equiv \cof{d}{00}/\sqrt{4\pi}$.
The special frame is often assumed to be the rest frame
of the cosmic microwave background radiation,
which differs from an Earth-based frame 
by a boost velocity $\simeq 10^{-3}$,
implying that the isotropic dispersion relation \rf{iso}
can provide only an approximation
to an exact treatment of Lorentz violation involving astrophysical neutrinos. 
Another disadvantage of the isotropic approximation
is that neutrino speeds exceeding light speed
occur only for negative $\ccfc{d}{}$,
so analyses based on threshold effects or \cv\ radiation
can yield only lower bounds on the isotropic coefficients.
In contrast,
the direction dependence encoded in the full dispersion relation \rf{dr}
permits two-sided bounds, 
given the availability of data with sufficient sky coverage.

While there is evidence that the IceCube TeV-PeV neutrinos 
are astrophysical 
\cite{icecube1,icecube2},
an atmospheric origin for these events cannot yet definitively be excluded.
We therefore begin by considering this more conservative scenario
for the two PeV IceCube events
\cite{icecube1}.

Suppose first that muon neutrinos or antineutrinos are produced
by atmospheric decays of hadrons $h$ via $h \rightarrow \mu + \nu_\mu$.
In the presence of unconventional dispersion relations,
these decays can display sharp threshold effects,
including becoming forbidden or allowed according to the neutrino energy 
\cite{ftlnu,cg99}.
Conservation of energy implies a threshold
on the energy defect
$\de E(\pvec) \equiv E(\pvec) - \pmag$
given by 
$\de E(\pvec) \leq \half \De M^2/\pmag$,
where $\De M = M_h - M_\mu$ is the difference 
between the hadron and muon masses
\cite{km12}.
Above this energy,
production of atmospheric neutrinos cannot occur.
The dispersion relation \rf{dr}
shows this effect is generically direction dependent,
but since only two PeV events are available
we work in the isotropic limit \rf{iso} for which
the threshold condition becomes 
\beq
-\sum_d \pmag^{d-2} \ccfc{d}{} \leq \half \De M^2.
\label{isokin}
\eeq
At lower energies,
the atmospheric neutrino flux is dominated by the decays 
of $\pi$ and $K$ mesons,
for which  
$\half\De M^2 \simeq 5.8\times 10^{-4}\text{ GeV}^2$
and
$\half\De M^2 \simeq 7.5\times 10^{-2}\text{ GeV}^2$,
respectively.
However, 
at higher energies
the prompt decays of short-lived charmed hadrons 
with $M \simeq 2$ GeV are expected to dominate,
for which $\half\De M^2 \simeq 2 \text{ GeV}^2$.
Taking one coefficient in Eq.\ \rf{isokin} at a time,
this value of $\De M^2$ and $\pmag \simeq 1$ PeV
yields estimated one-sided constraints on 
the isotropic SME coefficients $\ccfc{d}{}$ 
under the assumption of an atmospheric origin
for the two PeV events.
The results for dimensions $d\leq 10$ are 
compiled in the second column of Table \ref{iso_bounds}.

\begin{table}
\begin{tabular}{c|c|c|c}
& Atmospheric & Atmospheric & Astrophysical \\
Coefficient & threshold & \cv\ & \cv\ \\
\hline\hline
$\ccfc{4}{}$	
& $> -2 \times 10^{-12}$
& $> -3 \times 10^{-13}$ 
& $> -5 \times 10^{-19}$
\\
$\ccfc{6}{}$	
& $> -2 \times 10^{-24}$ 
& $> -3 \times 10^{-25}$
& $> -5 \times 10^{-31}$
\\
$\ccfc{8}{}$	
& $> -2 \times 10^{-36}$ 
& $> -2 \times 10^{-37}$ 
& $> -5 \times 10^{-43}$
\\
$\ccfc{10}{}$	
& $> -2 \times 10^{-48}$ 
& $> -2 \times 10^{-49}$ 
& $> -5 \times 10^{-55}$ 
\end{tabular}
\caption{\label{iso_bounds}
Estimated lower bounds 
on the isotropic coefficients $\ccfc{d}{}$
obtained using the two PeV IceCube events
\cite{icecube1}.
Units are GeV$^{4-d}$.}
\end{table}

Comparable sensitivities can be obtained
through limits on the energy loss 
due to \cv-like decays of atmospheric neutrinos
as they propagate to the detector.
The \cv\ radiation can occur when the maximum attainable neutrino speed 
exceeds that of the emitted particles,
which for the Lorentz-invariant case is the speed of light
\cite{cg}.
Consider a superluminal neutrino of 4-momentum $p=(E,\pvec)$
that decays via the neutral-current process
$\nu_\mu\rightarrow\nu_\mu + e^+ + e^-$
into a neutrino of momentum $p'=(E',\pvec')$ 
and a charged-lepton pair with momenta 
$k=(k^0,\kvec)$ and $k'=(k'^0,\kvec')$.
The rate of energy loss is given by an integral
of the form 
\cite{km12}
\beq
\frac{dE}{dx} = -\frac{C}{8} \int
\frac{\ka^0 \kavec'^2\tfrac{\prt|\kavec'|}{\prt\ka_0}}
{(\ka^2-M_Z^2)^2}
\frac{q\!\cdot\!k\ q'\!\cdot\!k'}
{q_0k_0q_0'k'_0} 
\ d^3p'\ d\Om_{\ka'} ,
\label{dEdx}
\eeq
where $C = 2 G_F^2 (1-4\sin^2\th_W + 8\sin^4\th_W)M_Z^4/(2\pi)^5$,
the 4-vectors $\ka$, $\ka'$ are defined as $\ka = k + k'$, $\ka' = k - k'$,
and $q/q^0 = (1, \phat)$, $q'/q'^0 = (1, \phat')$.
The solid angle $d\Om_{\ka'}$ is associated with the vector $\kavec'$,
while the integral is restricted to the phase space 
for which $p=p'+k+k'$.
A typical baseline distance for atmospheric neutrinos 
is on the order of 1000 km.
Neutrinos with significant superluminal speeds 
would dissipate much of their energy 
via \cv\ pair production before being detected,
so the observation of high-energy neutrinos 
implies a limit on the rate of energy loss.
The characteristic propagation distance associated 
with this energy loss is given by the
distortion distance $D(E) = -E/(dE/dx)$.
Numerically performing the integral in Eq.\ \rf{dEdx},
we can determine $D(E)$ for various fixed values 
of each coefficient $\ccfc{d}{}$ in turn.  
Large negative values of $\ccfc{d}{}$ produce a distortion distance 
much smaller than 1000 km,
which would imply a substantial energy loss.
Identifying the value of $\ccfc{d}{}$
for which $D(E)\simeq 1000$ km therefore
gives an estimated lower constraint on $\ccfc{d}{}$.
The third column of Table \ref{iso_bounds}
lists the bounds obtained in this manner for $d=4,6,8,10$,
assuming an atmospheric origin for the PeV IceCube events.

Significantly tighter constraints hold
if the high-energy neutrinos detected by IceCube are of astrophysical origin,
as has been suggested by several researchers
\cite{icecube1,icecube2,astronu}.
The large propagation distance implies 
that even a minuscule $dE/dx$ has a substantial effect.
Neutrinos above the threshold for \cv-like decays 
lose energy until they are at or near threshold,
so observed astrophysical neutrinos
must have energies near or below threshold.
For a positive energy defect $\de E$, 
the threshold energy is 
\bea
E(\pvec)
&=& \sqrt{\kvec^2+m_e^2} + \sqrt{\kvec'^2+m_e^2} + E(\pvec')
\nn\\
&\geq& \sqrt{(\kvec+\kvec')^2 + 4m_e^2} + \sqrt{\pvec'^2}
\nn\\
&\geq& 
\sqrt{\pvec^2 + 4m_e^2} .
\qquad
\eea
Squaring both sides of the above inequality
and dropping the small $\de E^2$ term 
produces the threshold condition 
$\pmag \de E(\pvec) \approx 2m_e^2$.
The condition that observed neutrinos 
are near or below this threshold then yields 
\beq
-\sum_{djm} \pmag^{d-2} Y_{jm}(\phat) \cof{d}{jm} \lsim 2m_e^2 .
\label{bound}
\eeq
In the isotropic limit and assuming an astrophysical origin,
this inequality implies the two PeV IceCube events
provide the lower bounds on $\ccfc{d}{}$
listed in the last column of Table \ref{iso_bounds},
where each coefficient is taken nonzero in turn.
For $d=4$,
these results are consistent with analyses of isotropic velocity defects
\cite{bcms}.
For each listed value of $d$,
the values obtained here sharpen by about a millionfold
the existing constraints on isotropic coefficients for Lorentz violation
in neutrinos,
reaching for the first time levels competitive 
with other SME astrophysical constraints 
from photon and fermion dispersion relation
\cite{km13}.

The above analysis shows that the two PeV events 
lead to stringent one-sided limits 
on simple isotropic models involving only the coefficients $\ccfc{d}{}$.
However,
the complete two-sided space of coefficients $\cof{d}{jm}$ for each $d$
is accessible only with a larger number of events 
involving neutrinos with different propagation directions.
For each observed event,
the inequality \rf{bound} provides a one-sided bound 
on a linear combination of the coefficients $\cof{d}{jm}$
fixed by the magnitude $\pmag$ and direction $\phat$
of the neutrino momentum,
representing a boundary plane 
in the $(d-1)^2$-dimensional coefficient space.
One can therefore expect that at least $(d-1)^2+1$ events 
are needed to extract two-sided constraints 
on all the possible types of Lorentz violation
allowed by operators of mass dimension $d$.
In practice,
the events must also be sufficiently well distributed
across the sky to insure their linear independence. 

Ideally,
combining constraints from multiple neutrinos 
would enclose a small volume in the coefficient space 
containing the zero-coefficient Lorentz-invariant limit.
Probing the allowed range of coefficients within this volume
would then permit the identification of robust two-sided limits
on individual coefficients for Lorentz violation.
However,
this ideal scenario is unattainable in practice 
because the direction-independent isotropic coefficient $\ccfc{d}{}$
enters all linear combinations of the form \rf{bound} 
accompanied by a negative multiplier,
and hence it has no upper bound.
At best, 
the bounding surface opens in the positive $\ccfc{d}{}$ direction,
so all the anisotropic coefficients are unconstrained 
as $\ccfc{d}{}\rightarrow\infty$.
Individual two-sided bounds are therefore impossible 
in the absence of a two-sided bound on $\ccfc{d}{}$.
Instead,
we determine here complete limits on anisotropic effects
for each $j\neq 0$
and then comment on the prospects for isotropic bounds.
We find that the 28 IceCube events at TeV-PeV energies 
\cite{icecube1,icecube2}
suffice in both number and sky distribution
to place complete constraints at $d=4$ and $d=6$ for each $j$.
Note that the existing data also permit
partial coverage of cases with $d\geq 8$,
but a complete treatment remains refractory 
until further events are accumulated. 

Numerical bounds can be calculated using 
a modified simplex method of linear programming \cite{lp},
which we briefly describe here in the context of 
the nine anisotropic coefficients with $d=4$
as an illustration.
The procedure begins by writing the 28 individual bounds
obtained from Eq.\ \rf{bound} in the form
\beq
-\sum_{jm} \Big[\pmag^2 Y_{jm}(\phat)/(2m_e^2 s)\Big] 
\Big(s \cof{4}{jm}\Big) 
< 1 ,
\eeq
where $s$ is a constant scaling factor
chosen so that the terms in square brackets are of order one,
thereby reducing precision error.
The nine scaled coefficients $s \cof{d}{jm}$ 
are placed in a column matrix $c$
with entries labeled $c_n$, $n=1,2,\ldots,9$.
The 28 constraints can then be written
as the matrix equation $A\cdot c < B$,
where $B$ is a 28-dimensional column matrix
and $A$ is a $28\times 9$ matrix of constants.
We can account for the possibility of unbounded coefficients 
by augmenting the 28 constraint equations 
by the conditions $c_n < \infty$ and $-c_n < \infty$,
where $\infty$ is the numerical infinity. 
This increases the row dimension of
our matrix inequality to $28+9+9=46$,
with the components of the matrix $B$ being either $1$ or $\infty$.
Adding these constraints allows the search method to step
to the boundary at infinity,
indicating an unbounded coefficient.
The simplex technique then introduces $46$ nonnegative slack variables 
in a column matrix $S$
and writes the matrix inequality as $A\cdot c + S = B$.
The procedure starts with the solution $c=0$, $S=B$
and then takes steps within the bounded region
to maximize a given coefficient.
The initial basic variables are the slack variables,
and the initial free variables are the coefficients,
so a tableau of the form 
\beq
\begin{array}{c|cccccccc|c}\small
& c_1 & c_2 & \ldots & c_9 & S_1 & S_2 & \ldots & S_{46} & \\
\hline
S_1 & A_{1,1}  & A_{1,2}  & \ldots & A_{1,9}  & 1 & 0 & \ldots & 0 & B_1 \\
S_2 & A_{2,1}  & A_{2,2}  & \ldots & A_{2,9}  & 0 & 1 & \ldots & 0 & B_2 \\
\vdots & \vdots   & & & & & & & \vdots & \vdots \\
S_{46} & A_{46,1} & A_{46,2} & \ldots & A_{46,9} & 0 & 0 & \ldots & 1 & B_{46}
\end{array}
\nn
\eeq
characterizes the initial system.
The first column lists the basic variables,
while the last column tracks their values.
The other columns are associated with 
the coefficients and the slack variables.
A standard matrix pivot operation moves a zero free variable 
into the list of nonzero basic variables,
displacing one of the original basic variables.
The strategy behind maximizing a coefficient $c_n$
has two steps: 
first, 
make $c_n$ a basic variable by performing 
a pivot around the element in the $c_n$ column
that gives the largest value of $c_n$;
and second,
repeatedly try all pivots leaving $c_n$ as a basic variable,
accepting those that increase $c_n$
and exiting when no allowed pivots increase $c_n$.
In both steps,
pivots that give negative slack variables are rejected.
The minimum for the coefficient $c_n$ can be found
by writing the matrix inequality as $(-A)\cdot(-c) < B$
and finding the maximum of $-c_n$,
which in practice means that
minima are obtained by changing the sign of the $A$ matrix
and applying the above maximization procedure.
Finally, 
the maxima and minima are divided by the scale factor $s$ 
to yield the desired bounds on the coefficient combinations.

\begin{table}
\begin{tabular}{cc||c|c|c}
$d$ & $j$ & Lower bound & Coefficient & Upper bound \\
\hline\hline
 4   &   0   &   $   -4  \times 10^{  -19  }  < $ & $      \cof{  4   }{  0   0   } $ &  \\ 
\hline
 4   &   1   &   $   -1 \times 10^{  -17  }  < $ & $      \cof{  4   }{  1   0   } $ & $ <  4  \times 10^{  -17  }  $   \\ 
     &       &   $   -3  \times 10^{  -17  }  < $ & $  \re \cof{  4   }{  1   1   } $ & $ <  2  \times 10^{  -17  }  $   \\ 
     &       &   $   -2  \times 10^{  -17  }  < $ & $  \im \cof{  4   }{  1   1   } $ & $ <  2  \times 10^{  -17  }  $   \\ 
\hline
 4   &   2   &   $   -1  \times 10^{  -17  }  < $ & $      \cof{  4   }{  2   0   } $ & $ <  7  \times 10^{  -17  }  $   \\ 
     &       &   $   -2  \times 10^{  -17  }  < $ & $  \re \cof{  4   }{  2   1   } $ & $ <  3  \times 10^{  -17  }  $   \\ 
     &       &   $   -2  \times 10^{  -17  }  < $ & $  \im \cof{  4   }{  2   1   } $ & $ <  5  \times 10^{  -17  }  $   \\ 
     &       &   $   -5  \times 10^{  -17  }  < $ & $  \re \cof{  4   }{  2   2   } $ & $ <  2  \times 10^{  -17  }  $   \\ 
     &       &   $   -3  \times 10^{  -17  }  < $ & $  \im \cof{  4   }{  2   2   } $ & $ <  4  \times 10^{  -17  }  $   \\ 
\hline
 6   &   0   &   $   -3  \times 10^{  -31  }  < $ & $      \cof{  6   }{  0   0   } $ &  \\
\hline
 6   &   1   &   $   -2  \times 10^{  -28  }  < $ & $      \cof{  6   }{  1   0   } $ & $ <  9  \times 10^{  -28  }  $   \\ 
     &       &   $   -6  \times 10^{  -28  }  < $ & $  \re \cof{  6   }{  1   1   } $ & $ <  5  \times 10^{  -28  }  $   \\ 
     &       &   $   -3  \times 10^{  -28  }  < $ & $  \im \cof{  6   }{  1   1   } $ & $ <  3  \times 10^{  -28  }  $   \\ 
\hline
 6   &   2   &   $   -4  \times 10^{  -28  }  < $ & $      \cof{  6   }{  2   0   } $ & $ <  7  \times 10^{  -27  }  $   \\ 
     &       &   $   -1  \times 10^{  -27  }  < $ & $  \re \cof{  6   }{  2   1   } $ & $ <  2  \times 10^{  -27  }  $   \\ 
     &       &   $   -1  \times 10^{  -27  }  < $ & $  \im \cof{  6   }{  2   1   } $ & $ <  3  \times 10^{  -27  }  $   \\ 
     &       &   $   -5  \times 10^{  -27  }  < $ & $  \re \cof{  6   }{  2   2   } $ & $ <  6  \times 10^{  -28  }  $   \\ 
     &       &   $   -1  \times 10^{  -27  }  < $ & $  \im \cof{  6   }{  2   2   } $ & $ <  4  \times 10^{  -27  }  $   \\ 
\hline
 6   &   3   &   $   -1  \times 10^{  -26  }  < $ & $      \cof{  6   }{  3   0   } $ & $ <  4  \times 10^{  -27  }  $   \\ 
     &       &   $   -2  \times 10^{  -27  }  < $ & $  \re \cof{  6   }{  3   1   } $ & $ <  1  \times 10^{  -26  }  $   \\ 
     &       &   $   -5  \times 10^{  -27  }  < $ & $  \im \cof{  6   }{  3   1   } $ & $ <  3  \times 10^{  -27  }  $   \\ 
     &       &   $   -2  \times 10^{  -27  }  < $ & $  \re \cof{  6   }{  3   2   } $ & $ <  1  \times 10^{  -26  }  $   \\ 
     &       &   $   -4  \times 10^{  -27  }  < $ & $  \im \cof{  6   }{  3   2   } $ & $ <  6  \times 10^{  -27  }  $   \\ 
     &       &   $   -5  \times 10^{  -27  }  < $ & $  \re \cof{  6   }{  3   3   } $ & $ <  6  \times 10^{  -27  }  $   \\ 
     &       &   $   -1  \times 10^{  -26  }  < $ & $  \im \cof{  6   }{  3   3   } $ & $ <  7  \times 10^{  -28  }  $   \\ 
\hline
 6   &   4   &   $   -5  \times 10^{  -27  }  < $ & $      \cof{  6   }{  4   0   } $ & $ <  2  \times 10^{  -27  }  $   \\ 
     &       &   $   -1  \times 10^{  -27  }  < $ & $  \re \cof{  6   }{  4   1   } $ & $ <  3  \times 10^{  -27  }  $   \\ 
     &       &   $   -1  \times 10^{  -27  }  < $ & $  \im \cof{  6   }{  4   1   } $ & $ <  6  \times 10^{  -28  }  $   \\ 
     &       &   $   -1  \times 10^{  -27  }  < $ & $  \re \cof{  6   }{  4   2   } $ & $ <  3  \times 10^{  -27  }  $   \\ 
     &       &   $   -2  \times 10^{  -27  }  < $ & $  \im \cof{  6   }{  4   2   } $ & $ <  1  \times 10^{  -27  }  $   \\ 
     &       &   $   -1  \times 10^{  -27  }  < $ & $  \re \cof{  6   }{  4   3   } $ & $ <  9  \times 10^{  -28  }  $   \\ 
     &       &   $   -2  \times 10^{  -27  }  < $ & $  \im \cof{  6   }{  4   3   } $ & $ <  1  \times 10^{  -27  }  $   \\ 
     &       &   $   -2  \times 10^{  -27  }  < $ & $  \re \cof{  6   }{  4   4   } $ & $ <  1  \times 10^{  -27  }  $   \\ 
     &       &   $   -5  \times 10^{  -28  }  < $ & $  \im \cof{  6   }{  4   4   } $ & $ <  1  \times 10^{  -27  }  $ 
\end{tabular}
\caption{\label{d4_aniso_bounds}
Constraints on dimensionless coefficients 
$\cof{4}{jm}$ and on $\cof{6}{jm}$ in GeV$^{-2}$
obtained using IceCube data 
\cite{icecube1,icecube2}.} 
\end{table}

Table \ref{d4_aniso_bounds} displays constraints 
on the SME coefficients $\cof{4}{jm}$ and $\cof{6}{jm}$ 
obtained using this simplex method
applied to the IceCube data 
\cite{icecube1,icecube2}
and reported in the Sun-centered frame
\cite{sunframe}.
We adopt a cosmological origin for the IceCube events,
but the order of magnitude of the constraints displayed holds 
even for a galactic origin.
For the complex coefficients with $m\neq 0$,
bounds on the real and imaginary parts are found separately.
The first column lists the values of $d$ and $j$ used in the routine,
while the third column contains the coefficient involved. 
The second and fourth columns provide the
resulting numerical lower and upper bounds, respectively.
For oscillation-free coefficients with $d=4$,
the results in Table \ref{d4_aniso_bounds}
represent improvements of well over a millionfold 
over existing constraints 
\cite{tables},
while for $d=6$ 
the improvements are by factors up to about $10^{20}$.

An interesting open issue is the prospects for 
independent {\it upper} bounds on $\ccfc{d}{}$.
For $d=4$, 
the current best upper bound is provided by Altschul 
\cite{ba},
who finds $\ccfc{4}{}\lsim 10^{-11}$ is required
to exclude the proton decay $p\to n + e^+ + \nu$ 
in cosmic rays with energies $\simeq 10^{20}$ eV. 
Generalizing this analysis to arbitrary $d$
and neglecting possible Lorentz violation in protons,
we obtain 
\beq
\ccfc{d}{} \lsim \fr{m_n}{\pmags{p}^{d-3}} 
\simeq 10^{33-11d} {\rm ~GeV}^{4-d}.
\label{cr}
\eeq
This yields $\ccfc{6}{}\lsim 10^{-33}$ GeV$^{-2}$ 
and offers good prospects for constraints on $\ccfc{d}{}$ for $d\geq 8$.
However,
achieving a competitive upper bound on $\ccfc{4}{}$ is challenging.
An interesting option is time-of-flight measurements,
which are sensitive to $\ccfc{4}{}$
because the time delay $\De t$ in a neutrino pulse of energy $\pmag$
arriving from a source at distance $L$ is
$\De t \approx L \pmag^{d-4} \ccfc{d}{}$
\cite{km12}.
For $d\geq 6$ this result is typically less sensitive than 
the constraint \rf{cr} from cosmic rays,
but for $d=4$ it is independent of energy
and offers interesting prospects.
For example,
time delays of 10 s or better could be observable
in future neutrino-photon coincidence measurements 
from a gamma-ray burst at a Gpc distance,
which would yield the constraint 
$\ccfc{4}{} \lsim 10^{-16}$ for $d=4$
along with the bounds
$\ccfc{d}{} \lsim 10^{8-6d}$ GeV$^{4-d}$ 
for arbitrary $d$ assuming
an accompanying emission of high-energy PeV neutrinos.
Another option could be neutrino pulse dispersion,
which produces a velocity difference $\de v$
between neutrinos of energies $\pmags 1$ and $\pmags 2$
given by 
\cite{km12}
$\de v = |(d-3) (\pmags{2}^{d-4}- \pmags{1}^{d-4}) \ccfc{d}{}|$.
However,
for $d=4$ this vanishes,
while for $d\geq 6$ 
the resulting constraints are weaker 
than the cosmic-ray bound \rf{cr}.

In this work,
we have demonstrated that observations 
of high-energy astrophysical neutrinos
place stringent limits on deviations from the laws of relativity.
The first IceCube events already 
improve the  constraints on Lorentz violation 
by factors ranging from about a million to about $10^{20}$,
making them competitive with other extreme astrophysical limits 
from photons and fermions.
The sensitivities achieved have surpassed
the level at which Planck-suppressed effects
could be expected to emerge,
and as such they place tight constraints on models
involving Planck-suppressed Lorentz violation.
Future observations in this new arena of astrophysics
offer excellent prospects for further sharpening
constraints on the available coefficient space.

This work was supported in part
by the Department of Energy
under grant DE-FG02-13ER42002
and by the Indiana University Center for Spacetime Symmetries.

\end{document}